\documentstyle[epsf]{elsart}  

\begin{document} 

\hfill \vbox{\hbox{CRN 96--01} \hbox{TK 96 02} \hbox{hep-ph/9601nnn}}

\bigskip \bigskip

\begin{frontmatter} 
\title{IMPROVED DESCRIPTION  OF THRESHOLD PION ELECTROPRODUCTION IN 
       CHIRAL PERTURBATION THEORY} 


\author[Strasbourg]{V.~Bernard}\footnote{email: 
bernard@crnhp4.in2p3.fr},
\author[TUM]{N.~Kaiser}\footnote{email: nkaiser@physik.tu-muenchen.de},
\author[Bonn]{Ulf-G.~Mei\ss ner}\footnote{email: 
                                          meissner@pythia.itkp.uni-bonn.de}


\address[Strasbourg]{Laboratoire de Physique Th\'eorique,
Institut de Physique \\ 3-5 rue de l'Universit\'e, F-67084 Strasbourg
Cedex, France\\
Centre de Recherches Nucl\'eaires, Physique Th\'eorique\\ 
BP 28, F-67037 Strasbourg Cedex 2, France}
\address[TUM]{Technische Universit\"at M\"unchen, Physik Department T39\\
James-Franck-Stra\ss e, D-85747 Garching, Germany}
\address[Bonn]{Institut f\"ur Theoretische Kernphysik, Universit\"at
  Bonn\\ Nu\ss allee 14--16, D-53115 Bonn,  Germany}


\begin{abstract}
We investigate neutral pion electroproduction off protons in the
framework of heavy baryon chiral perturbation theory. The chiral
expansion of the S--wave multipoles $E_{0+}$ and $L_{0+}$ is carried
out to three orders. There appear several undetermined low--energy
constants. 
Three are taken from a recent study of the new TAPS threshold 
$\pi^0$ photoproduction data, one is fixed from 
the proton Dirac  form factor and 
the  novel S--wave constants appearing at
orders $q^4$ and $q^5$ are determined from a best 
fit (constrained by resonance exchange) to the 
recent NIKHEF and MAMI data at photon momentum transfer 
squared $k^2 = -0.1 \, \,  {\rm GeV}^2$.  
The inclusion of a particular set of dimension five operators is
forced upon the fit by a soft--pion theorem which severely constrains
the momentum dependence of the longitudinal S--wave multipole $L_{0+}$
at order $q^4$. We give predictions for lower photon virtualities for
the various multipoles and differential cross sections.  Further
improvements are briefly touched upon.
\end{abstract}

\end{frontmatter}
 


\section{Introduction}

Chiral perturbation theory (CHPT) allows to systematically investigate the
consequences of the spontaneous and the explicit chiral symmetry
breaking in QCD. With the advent of CW machines, pion production by 
real and virtual photons has become a major testing ground for
predictions based on nucleon chiral dynamics. In particular,
over the last years there has been considerable
activity to precisely measure neutral pion electroproduction in the
threshold region at various laboratories, in particular at NIKHEF
(Amsterdam) and MAMI (Mainz). These measurements are performed at
energies very close to threshold and typical photon momentum transfer
squared of $k^2 =-0.1 \, \, {\rm GeV}^2$. After the pioneering
measurement of Welch et al. \cite{pat} at NIKHEF, which concentrated 
on the S--wave cross section, kinematically more complete differential
cross sections for  various photon polarizations $\epsilon = 0.5
\ldots 0.9$ are now available based on the data of van den Brink et
al. \cite{benno} (NIKHEF) and Distler et al. \cite{dist}
(MAMI). Furthermore, data at lower $|k^2|$ have also been taken at MAMI
\cite{dwpc}.

In the review \cite{bklm} we already considered charged and neutral
pion electroproduction off protons and neutrons in the framework of
relativistic nucleon CHPT \cite{gss} to order $q^3$ (here, $q$ denotes
the small expansion parameter which can be an external momentum or
meson mass and the order $q^n$ refers to the Lagrangian). 
Since then, the theoretical framework has been
considerably refined based on the heavy mass expansion proposed in
\cite{jm} and discussed in detail in the recent review \cite{bkmr}. In
this framework and calculating the S--waves to order $q^4$, we have
already reexamined neutral pion photoproduction
\cite{bkmz,bkmprl}. Novel P--wave low--energy theorems (LETs) have been found
together with a good description of the new TAPS and SAL data 
for $\gamma p \to \pi^0 p$ \cite{fuchs} \cite{pi0sal}. 
In particular, chiral loops are necessary to understand the small value
of the electric dipole amplitude $E_{0+}$ at the $\pi^0 p$ and $\pi^+ n$
thresholds. Furthermore, the P--wave LETs have been shown to hold
within 10\% or better. While $P_1$ can be inferred directly from the
unpolarized differential cross section, 
for $P_2$ one has to measure polarization 
observables. A measurement of the photon polarization asymmetry is
 underway e.g. at MAMI. However, there are somewhat 
model--dependent means to indirectly determine $P_2$, see e.g.
\cite{berg} and \cite{olaf}. These LET predictions extended to the
case of virtual photons \cite{bkmlet} were also used in 
the analysis of the NIKHEF data \cite{benno}. The investigation
presented here is thus a natural extension of the one for
photoproduction presented in \cite{bkmz}. To orders $q^3$ and $q^4$, 
there appear two new counter terms with a priori undetermined
low--energy constants (LECs). One can be fixed from a recent
dispersion--theoretical fit to the nucleon electromagnetic form
factors, i.e. from the radius of the proton Dirac form factor \cite{mmd}. 
The other one will be determined from a best fit to the
differential cross sections at $k^2 = -0.1 \, \, {\rm GeV}^2$.
As we will show, due to a soft--pion theorem, these fits are too
constrained since they lead to a $k^2$--dependence of the longitudinal S--wave
multipole $L_{0+}$ only through Born and one--loop graphs. We therefore are
forced to include the leading corrections to this soft--pion theorem
in the counter terms which is formally of order $q^5$ in the
Lagrangian. This in turn
leads to a satisfactory fit of the existing data at $k^2 =
-0.1$~GeV$^2$  and we then make predictions for smaller values of
$|k^2|$, where this dimension five operator is much less important.
In addition, three $k^2$--independent low--energy constants are taken
from the photoproduction calculation to order $q^4$ \cite{bkmprl}.

The paper is organized as follows. In section 2, we briefly summarize
the pertinent kinematics and define the various cross sections and
structure functions. We also present the natural basis of S-- and
P--wave multipoles which appear in the transition matrix--element.
In section 3, the chiral expansion of the S--wave multipoles $E_{0+}$
and $L_{0+}$ is given to three orders in small momenta and the five
combinations of P--wave multipoles to two orders.
The corresponding low--energy theorems were already given 
in \cite{bkmlet}.  We also introduce an
approximation which facilitates the calculation of the
energy--dependence of the S--wave multipoles. In section 4, we give
the explicit expression for the various low--energy constants based on
the resonance saturation  principle. In section 5, the best fits to the
differential cross sections are presented together with a detailed
analysis of the various multipoles at $k^2 = -0.1 \, \, {\rm GeV}^2$.
In section 6, we give predictions for the new data taken at $k^2 = -0.06 \, \, 
{\rm GeV}^2$ and investigate in some detail the $|k^2|$ range of $0.04
\ldots 0.06$~GeV$^2$.  A short summary together with a discussion of 
possible improvements is given in section 7. The appendix contains some
consideration about the S--wave LECs at order $q^4$.


\section{Formal aspects}

In this section, we assemble all the necessary definitions for the
cross sections, structure functions and multipoles. To be specific,
consider the process
\begin{equation}
\gamma^\star \, (k) + p \, (p_1) \to \pi^0 \, (q) + p \, (p_2) \, \, ,
\end{equation}
where $\gamma^\star$ is the virtual photon with $k^2 <0$. 
Denote by $W$ the cm energy of the $\pi N$ system and
its threshold value by $W_0 = M_{\pi} + m$, with $M_\pi =134.97 \,
\, {\rm MeV}$ and $m =938.27 \, \, {\rm MeV}$ the (neutral) pion and
the proton mass, respectively. 
The following quantities are used
\begin{eqnarray} 
q &=& \sqrt{ \omega^2 - M_{\pi}^2} \, \, , \quad 
k_0 = \frac{1}{2 W} (W^2 -m^2 +k^2) \, \, ,\nonumber \\  
\epsilon^{-1} &= & 1 +2 \bigl(1- \frac{k_0^2}{k^2} \bigr)\, 
\tan^2 \frac{\psi}{2} \, \, , \quad
\epsilon_L = - \frac{k^2}{k_0^2 }\, \epsilon \, \, , \nonumber \\ 
\omega &=& \frac{1}{2W}(W^2  -m^2 + M_\pi^2 ) \, \, , \quad
\Delta W = W - W_0 \, \, , 
\label{kine1}
\end{eqnarray}
with $q$ the pion cm momentum, $k_0$ the photon energy, $\epsilon$
the photon polarization, $\epsilon_L$ the longitudinal photon
polarization, $\psi$ the electron scattering angle, $\omega$ the
pion energy in the $\pi N$ cm system and $\Delta W$ gives the invariant
energy above threshold. 
As explained in some detail in \cite{bkmz}, we will account for the
pion mass difference $M_{\pi^+} - M_{\pi^0}$ in the loops but keeping
one nucleon mass. The $\pi^+ n$ threshold is located at $\omega_c =
140.11 \, \, {\rm MeV}$ which will also be used as the charged pion
mass. We introduce the ratio
\begin{equation}
\rho = -\frac{k^2}{\omega_c^2} > 0 \quad ,
\label{defrho}
\end{equation}
and as a further dimensionless quantity of order one  
\begin{equation}
y = \frac{\omega^2}{\omega_c^2}  \quad .
\label{defy}
\end{equation}
$y$ varies between 0.93 and 1.12 for $\Delta W = 0 \ldots 15$ MeV.

The unpolarized pion electroproduction triple differential cross section reads
\begin{equation}
\frac{d\sigma}{ d E_f d\Omega_f d\Omega_\pi} = \frac{\alpha E_f (W^2-m^2)}{ 4 
\pi^2 E_i  m  k^2 ( \epsilon -1) } \, \frac{d\sigma }{ d\Omega_\pi} =
\Gamma_V \, \frac{d\sigma }{ d\Omega_\pi} \, \, ,
\label{3diff}
\end{equation}
with $\Gamma_V$ the conventional virtual photon flux factor, $\alpha =
e^2/4 \pi = 1/137.036$ the fine structure constant and $E_{i/f}$
is the laboratory energy of the incoming/outgoing electron.
The differential cross section can be split into transverse ($T$),
longitudinal ($L$), transverse--longitudinal ($TL$) and transverse-transverse
($TT$) terms,
\begin{eqnarray}
\frac{d \sigma}{ d\Omega_\pi} & =& \frac{2 W q}{ W^2-m^2}\biggl( R_T +
\epsilon_L\, R_L + \sqrt{2\epsilon_L(1+\epsilon)} \cos \phi \, R_{TL} +
\epsilon \cos 2 \phi \, R_{TT} \biggr)  , \nonumber \\
& &
\label{diffxs}
\end{eqnarray}
where the $R_I$ ($I=T,L,TL,TT$) are called the structure
functions. $\phi$ is the azimuthal angle between the scattering and
the reaction plane, and the corresponding polar angle,
spanned by the photon and pion directions, is called $\theta$,
\begin{equation}
\cos\theta = \hat q \cdot \hat k \quad .
\end{equation}  
One also uses the  separated virtual photon cross sections 
\begin{equation}
\frac{d \sigma_I }{ d \Omega_\pi} = {2W q \over W^2 - m^2} \, R_{I}, 
\quad I = T,L,TL,TT \quad .
\label{sepxs}
\end{equation}
This completes the necessary definitions of kinematical quantities and
cross sections.

In what follows,  we will consider the threshold region, i.e. small
values of $\Delta W$, typically $\Delta W < 15 \, {\rm MeV}$ and small photon 
virtualities, $|k^2| \le 0.1\, \, {\rm GeV}^2$. In that case, the pion
three momentum is small and it is therefore advantageous to perform a
multipole decomposition. We confine ourselves to the S-- and P--waves
in what follows. Consequently, the current matrix element takes the form
\begin{eqnarray}
{m \over 4 \pi W} \, \vec J & = & i \vec \sigma \, \biggl(E_{0+} + 
\hat q \cdot \hat k
\, P_1 \biggr) + i \vec \sigma \cdot \hat k \, \hat q \, 
P_2 + \hat q \times \hat k \, P_3  \nonumber \\ 
& + & i \vec \sigma \cdot \hat k \, \hat k\, \biggl(L_{0+} - E_{0+} +
\hat q \cdot \hat k\,(P_4 -P_5-P_1-P_2) \biggr) 
+ i  \vec \sigma \cdot \hat q \, \hat k \, P_5 \,  , \nonumber \\ & &
\label{defJ}
\end{eqnarray}
in terms of the two S--waves $E_{0+}$and $L_{0+}$ and five P--waves.
We choose
the following combinations of the more commonly used P--wave
multipoles $E_{1+}$, $M_{1\pm}$ and $L_{1\pm}$, where $E,M,L$ stands
for electric, magnetic and longitudinal, respectively and the $\pm$
refers to the total angular momentum of the pion--nucleon system, $j =
l \pm 1/2$ (with $l$ the pion angular momentum),
\begin{eqnarray}
 P_1 &=& 3 E_{1+}+M_{1+}-M_{1-}, \quad P_2 = 3 
E_{1+}-M_{1+}+M_{1-}, \nonumber \\
P_3 &=& 2M_{1+}+M_{1-}, \quad P_4 = 4 L_{1+}+L_{1-},
 \quad P_5 = L_{1-}-2 L_{1+} \, \, .
\label{Pi}
\end{eqnarray}
These are the combinations which appear naturally in the transition
matrix element and lead to the most compact formulae for the various
cross sections. 
All multipoles are, of course, functions of the pion
energy and the photon four--momentum squared, like e.g. $E_{0+} =
E_{0+} (\omega, k^2)$. In what follows, we will drop these obvious arguments.
The structure functions $R_I$ can be expressed in terms of the
multipoles as follows (in the S-- and P--wave approximation)
\begin{eqnarray}
R_T & = & |E_{0+}+ \cos \theta \, P_1|^2 + \frac{1}{ 2} \sin^2\theta\,
 ( |P_2|^2+|P_3|^2) \, , \nonumber \\  
R_L & = & |L_{0+}+ \cos \theta \, P_4|^2 + \sin^2\theta
\,|P_5|^2 \, , \nonumber \\  
R_{TL} & =& - \sin\theta\, {\rm Re}\bigl[ (E_{0+}+ \cos\theta \, 
P_1)\,P_5^* + (L_{0+}+ \cos\theta \,P_4)\, P_2^*\bigr] \, , \nonumber \\
R_{TT} & = & \frac{1}{2} \sin^2 \theta\, ( |P_2|^2 - |P_3|^2) \, \, .
\label{Rmult}
\end{eqnarray} 


\section{Chiral expansion of the multipoles}

To perform the calculations, we make use of the effective
Goldstone boson--baryon Lagrangian. Our notation is identical to the one
used in \cite{bkmz} and we discuss here only some additional terms.
The  effective Lagrangian takes the form
(to one loop accuracy)
\begin{equation}
{\cal L}_{\rm eff} = {\cal L}_{\pi N}^{(1)} +  {\cal L}_{\pi N}^{(2)} +
  {\cal L}_{\pi N}^{(3)} +
  {\cal L}_{\pi N}^{(4)} + {\cal L}_{\pi \pi}^{(2)}+ {\cal L}_{\pi \pi}^{(4)}
\label{leff}
\end{equation}
where the chiral dimension $(i)$ counts the number of derivatives
and/or meson mass insertions. We will work out the S--wave multipoles
to order $q^3$ and the P--waves to order $q^2$ as explained in some
detail in ref.\cite{bkmz}. Here, the order refers to the various
multipoles. The corresponding Lagrangians are one order higher since
the photon polarization vector counts as order $q$.  
For the P--waves, we thus have to consider
tree graphs with insertions from ${\cal L}_{\pi N}^{(1,2,3)}$ and one
loop graphs with insertions solely from $ {\cal L}_{\pi N}^{(1)}$. For
the S--wave multipoles, we have in addition to consider one loop
diagrams with exactly one insertion from ${\cal L}_{\pi N}^{(2)}$ and
tree graphs from ${\cal L}_{\pi N}^{(4)}$.
In comparison to the photoproduction calculation,
we have two additional terms with undetermined low-energy
constants. One is from ${\cal L}_{\pi N}^{(3)}$ and is related to the
Dirac form factor of the proton, $F_1^V (k^2)$, and was already
considered in the relativistic calculation in \cite{bklm}. In the
heavy fermion approach used here, the effect due to the finite charge
radius of the proton manifests itself via
\begin{equation}
<r^2>_1^p = \frac{1}{ 16 \pi^2 F_\pi^2 } \biggl[ -(5g_A^2+1) \ln 
\frac{\omega_c}{ \lambda}  - \frac{7}{2} g_A^2 -
\frac{1}{2} \biggr] + \delta r_{1p}(\lambda) \, \, \, ,
\label{F1v}
\end{equation}
with $\lambda$ the scale of dimensional regularization,
$g_A$ the axial--vector coupling constant and $F_\pi = 93 \, {\rm MeV}
$ the pion decay constant. The terms in the square brackets in
Eq.(\ref{F1v}) stem from the loops. The constant 
$\delta r_{1p}(\lambda)$ can be fixed from the recent
dispersion--theoretical fit to the nucleon electromagnetic form
factors, $<r^2>_1^{p, \exp} = 0.774 \pm 0.008 \, {\rm fm}^2$ \cite{mmd}.
The other counter term is from ${\cal L}_{\pi N}^{(4)}$ and
appears in the chiral expansion of the S--wave multipoles. We do not
need the explicit form of this term in the effective Lagrangian in
what follows and thus refrain from giving it here.\footnote{As
discussed later, we also need to take one particular term from 
${\cal L}_{\pi N}^{(5)}$ since with the sole 
$k^2$--dependent counter term from ${\cal L}_{\pi N}^{(4)}$ 
togther with the radius correction from ${\cal L}_{\pi N}^{(3)}$ 
one is not able to describe the existing data at rather large $|k^2|
\simeq 5 M_\pi^2$.}

Consider first the chiral expansion of the P--wave multipoles to order
$q^3$. This should be rather accurate for the large multipoles but is
afflicted with some uncertainty for the small ones, compare e.g. the
discussion by Bergstrom \cite{berg} for the photoproduction case. He
finds a good description of the large multipole $M_{1+} -M_{1-}$ but
some sizeable deviation for the much smaller $E_{1+}$. As we will show
later on, the existing electroproduction data are not yet accurate enough
to pin down the small multipoles with great precision and we thus
stick to the $q^3$ approximation.\footnote{This should eventually be
  refined when more accurate data will become available.} The chiral
expansion of the P-wave multipoles thus takes the form
\begin{equation}
P_i = P_i^{\rm Born} + P_i^{\rm loop} + P_i^{\rm ct} \, \, ,
 \quad i= 1,\ldots ,5 \, \, \, ,
\label{pigen}
\end{equation}
with $P_3^{\rm loop}=0$ and $P_i^{\rm ct} =0$ for $i = 1,2,4,5$.
The Born terms include the contribution from the proton anomalous
magnetic moment $\kappa_p$, which to lowest order
appears in the dimension two pion--nucleon Lagrangian.
 For the large multipoles $P_{1,2,3}$ we have
the following Born (and counter) terms (in case of $P_3$),
\begin{eqnarray}
P_1^{\rm Born} & = & \frac{e g_{\pi N}\,q}{ 8 \pi m^2\omega }
\biggl\{ (1 + \kappa_p ) \sqrt{\omega^2-k^2} + \frac{1}{ 10 m \omega
\sqrt{\omega^2 - k^2}} \times   \\
 & \bigl[ & 17 \omega^2 k^2-12 \omega^4  + 2
\omega^2 M_\pi^2 - 7 k^2 M_\pi^2 + 5 \kappa_p(2 \omega^2 k^2 -k^2 M_\pi^2 -
\omega^4) \bigr] \biggr\}  \, , \nonumber \\ & & \nonumber 
\label{p1born}
\end{eqnarray}

\vspace{-1.0cm}

\begin{eqnarray}
P_2^{\rm Born} & = & \frac{e g_{\pi N}\,q}{ 8 \pi m^2\omega}
\biggl\{-(1 + \kappa_p ) \sqrt{\omega^2-k^2} + \frac{1}{ 10 m \omega
\sqrt{\omega^2 - k^2}}  \times \\ 
&  \bigl[ & 13 \omega^4 -18 \omega^2 k^2 + 2
\omega^2 M_\pi^2 +3 k^2 M_\pi^2 + 5 \kappa_p(\omega^4 +k^2 M_\pi^2 -2\omega^2
k^2)  \bigr] \biggr\} \, , \nonumber \\ & & \nonumber 
\label{p2born}
\end{eqnarray}
\begin{equation}
P_3^{\rm Born+ct} = e\, q\biggl( \frac{g_{\pi N}}{16 \pi m^3 } + b_P\biggr)
\sqrt{\omega^2-k^2} \quad ,
\label{p3born}
\end{equation}
with $g_{\pi N} =13.4$ the strong pion--nucleon coupling constant. We
notice that the P--waves scale with the pion momentum $q$. The constant
$b_P$ has been determined from a best fit to the new TAPS data for
$\gamma p \to \pi^0 p$, $b_P = 13.0 \, {\rm GeV}^{-3}$, a value 
which is well understood in terms of $\Delta$ and vector meson
exchange. However, there
are also new data from SAL for the same process. The 
analysis of these  gives a somewhat higher total cross section
above $\pi^+ n$ threshold \cite{pi0sal}. To account for this, we
also use   the larger value of $b_P = 15.8 \, {\rm GeV}^{-3}$ as
determined in \cite{bkmz}. The small multipoles $P_{4,5}$ have the
following Born terms 
\begin{eqnarray}
P_4^{\rm Born} &=& \frac{ e g_{\pi N}\,q }{ 40 \pi m^3 } \biggl( 2 + 3
\frac{M_\pi^2 }{ \omega^2 } \biggr) \sqrt{\omega^2 - k^2} \, \, , \\
P_5^{\rm Born}& =& \frac{ e g_{\pi N} \,q}{ 80 \pi m^3 } \biggl( 3 + 2
\frac{M_\pi^2 }{ \omega^2 } \biggr) \sqrt{\omega^2 - k^2} \, \, .
\label{p45born}
\end{eqnarray}
All these Born terms are, of course, real. The one loop contribution to
the $P_i$ $(i=1,2,4,5)$ takes the form
\begin{eqnarray}
 P_1^{\rm loop} & = & \frac{e g_{\pi N}^3 \omega_c\,q }{ 32 \pi^2 m^3} \frac{1
}{ \sqrt{y(y+\rho)^3} } \biggl[ \frac{\rho}{3} +\frac{3}{8} \rho^2 +\frac{y}{3}
+ \frac{3}{4 }\rho\, y + \frac{y^2 }{2} -\sqrt{1-y} \nonumber \\
 & \times &  \bigl( \frac{\rho}{ 3}
+\frac{\rho^2}{ 8} + \frac{y}{3} +\frac{\rho\,y}{ 6} + \frac{y^2}{6} \bigr) -
\frac{(\rho+2y)(4\rho+\rho^2+4y)}{ 16 \sqrt{y+\rho}} H(y,\rho) \biggr]
\, \, , 
\label{p1loop}
\end{eqnarray}

\vspace{-1cm}

\begin{eqnarray}
P_2^{\rm loop} & = & \frac{e g_{\pi N}^3 \omega_c\,q }{32 \pi^2 m^3} \frac{1
}{ \sqrt{y(y+\rho)^3} } \biggl[ \frac{\rho}{3} -\frac{3}{8} \rho^2 +\frac{y}{3}
-\rho\, y - \frac{y^2}{2} + \sqrt{1-y} \nonumber \\
& \times & \bigl(- \frac{\rho}{3}
+\frac{\rho^2}{8} - \frac{y}{3} +\frac{7}{12} \rho\,y + \frac{y^2}{3} \bigr) +
\frac{\rho(4\rho+\rho^2+4y)}{16 \sqrt{y+\rho}} H(y,\rho) \biggr] 
\, \, ,
\label{p2loop}
\end{eqnarray}

\vspace{-1cm}

\begin{eqnarray}
P_4^{\rm loop} & = & \frac{e g_{\pi N}^3 \omega_c\,q}{32 \pi^2 m^3} \frac{1
}{\sqrt{y(y+\rho)^3} } \biggl[\frac{\rho}{3}+\frac{y}{3} -\frac{\rho\, y}{4} -
\frac{y^2}{2} +\frac{1}{3}\sqrt{1-y} \nonumber \\
& \times & \bigl(- \rho- y+\frac{\rho\,y}{ 4} +
y^2 \bigr)- \frac{(\rho+2y)\rho \,y}{8\sqrt{y+\rho}} H(y,\rho) \biggr]
\, \, , \label{p4loop}
\end{eqnarray}

\vspace{-1cm}

\begin{eqnarray}
P_5^{\rm loop} &= &  \frac{e g_{\pi N}^3 \omega_c\,q}{32 \pi^2 m^3} \frac{1
}{\sqrt{y(y+\rho)^3} } \biggl[ \frac{\rho}{3} +\frac{y}{3}-\frac{\rho\, y}{8}
 +\frac{1}{3}  \sqrt{1-y} \nonumber \\
& \times & \bigl( -\rho- y+\frac{5}{ 8} \rho\,y+ \frac{y^2}{ 4}
\bigr) - \frac{y(4\rho+\rho^2+4y)}{16 \sqrt{y+\rho}} H(y,\rho) \biggr]
\, \, , \label{p5loop}
\end{eqnarray}
with
\begin{equation}
H(y,\rho) = \arctan\frac{\rho}{2 \sqrt{y+\rho}} + \arcsin\frac{\rho+2y}
{\sqrt{4\rho+\rho^2+4y}} \, \,  , \, \, \,\, \omega < \omega_c \, \, .
\label{Hdef}
\end{equation}
Loop effects which renormalize physical quantities like e.g. the anomalous 
magnetic moment $\kappa_p$ or the pion--nucleon coupling
are properly taken care of.
For $\omega > \omega_c$, these loop contributions become
complex. This has to be accounted for by the substitutions
\begin{equation}
\sqrt{1-y}=-i \sqrt{y-1} \quad ,
\label{wurzi}
\end{equation}
and
\begin{equation}
\arcsin\frac{\rho+2y}{\sqrt{4\rho+\rho^2+4y}}= \frac{\pi}{2} + i\, 
\ln \frac{\rho+2y +2\sqrt{(y+\rho)(y-1)}}{\sqrt{4\rho+\rho^2 +4y}} \,
\, .
\label{Hi}
\end{equation}
For the imaginary parts of the P--wave multipoles, one can give a
rather compact and accurate description based on the Fermi--Watson
final state theorem. This serves as a consistency check and gives a
handy and rather accurate estimate about the size of the Im~$P_i$.  
Of course, in the full calculations to be performed later, we are not
using these approximate forms. Consider the rescattering diagram
shown in Fig.1. The corresponding imaginary part can be cast into
\begin{figure}[b]
\hskip 1.5in
\epsfysize=1.5in
\epsffile{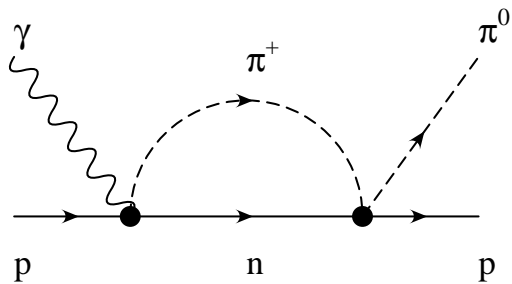}
\caption{\label{Fig. 1} Rescattering diagram. The solid circles
  subsume various subdiagrams.}
\end{figure}
the product of a charged pion production process followed by a
pion--nucleon charge exchange (CEX) reaction. Consequently
\begin{equation}
{\rm Im} \, ( {\cal M})^{\pi^0 p} = q_+^3 \cdot a_{2J}^{\rm CEX}
\, \, {\rm Re} ({\cal M})^{\pi^+ n} \, \, ,
\label{pwatson}
\end{equation}
where ${\cal M}$ is a generic symbol for any one of the P--wave
multipoles $E_{1+}$, $M_{1\pm}$, $L_{1\pm}$ and $J = 3/2 \, \, (1/2)$
for the $1+$ ($1-$) multipoles, $a_{2J}$ denotes the 
corresponding charge--exchange (CEX) scattering volume and 
$q_+ = \sqrt{\omega^2 - \omega^2_c}$. To lowest
order, these scattering volumina are given by
\begin{equation}
a_3^{\rm CEX} = a_1^{\rm CEX} = - \frac{ \sqrt{2} \,g_A^2}
{24 \pi M_\pi F_\pi^2} \, \, \, .
\label{a13cex}
\end{equation}
Furthermore, the P--waves for $\gamma p \to \pi^+ n$ can be easily
evaluated in Born approximation from the pion pole diagram,
\begin{equation}
{\vec J \,}^{\pi^+ n} = i \frac{e g_A}{\sqrt{2}F_\pi} \vec
\sigma \cdot ( \vec q - \vec k \, ) \, ( \vec k - 2
\vec q\, ) 
\, \biggl[  \frac{1}{(2+\rho) M_\pi^2} + \frac{ 2 \vec q \cdot \vec
  k }{(2+\rho)^2 M_\pi^4} \biggr] \, \, \, ,
\label{pi+}
\end{equation}
where the  terms in the square brackets come from the expansion of the
pion propagator. Evaluating the operator structure of Eq.(\ref{pi+})
 fixes the $P_i^{\pi^+ n}$. Combining these with
CEX scattering volumina and the appropriate kinematical factors, we
have
\begin{eqnarray}
{\rm Im} \, P_1^{\pi^0 p} &=& 0 \,\,\, , \, \, \, 
{\rm Im} \, P_2^{\pi^0 p} = -\frac{e \, g_A^3 \,  q \, q_+^3}{48 \,
\pi^2 \,  F_\pi^3 \,
  M_\pi^2} \frac{\sqrt{1+\rho}}{2+\rho}  \,\,\, , \nonumber \\
{\rm Im} \, P_4^{\pi^0 p} &=& \frac{4+\rho}{2(2+\rho)} \, 
{\rm Im} \, P_2^{\pi^0 p} \, \, \, , \, \, \, 
{\rm Im} \, P_5^{\pi^0 p} = \frac{1}{2} \, {\rm Im} \, P_2^{\pi^0 p}
\, \, . \label{appimpi}
\end{eqnarray}
This shows that $P_2$ has the largest imaginary part. These approximations
become very accurate as $\rho$ approaches zero. Isospin breaking
effects not taken into account in such a consideration are of the
order $(\omega_c^2 - M_{\pi^0}^2 )^{3/2}/M_{\pi^0}^3 \sim 2 \, \%$.

We now turn to the S--wave multipoles $E_{0+}$ and $L_{0+}$. Let
${\cal S}$ be a generic symbol for either one of them. The chiral
expansion carried out to order $q^4$ takes the form
\begin{equation}
{\cal S} = {\cal S}^{\rm Born} + {\cal S}^{\rm q^3-loop} 
+ {\cal S}^{\rm q^4-loop} + {\cal S}^{\rm ct} \, \, .
\label{s0+gen}
\end{equation}
As already mentioned, ${\cal S} = {\cal S} \, (\omega,
k^2)$. However, from our photoproduction study \cite{bkmz}, we know
that in the threshold region, $M_{\pi^0} < \omega < 150$ MeV,
 the $\omega$--dependence is mostly due to
the unitarity cusp which is fixed by the Fermi--Watson final state
theorem. Stated differently, this accounts for the dominant isospin
breaking effect due to the sizeable pion mass difference, $\sqrt{(M_{\pi^+}
-M_{\pi^0} )/ M_{\pi^0}} = 18\%$. We will discuss the numerical 
implications of this approximation later on. Consequently 
\begin{eqnarray}
E_{0+}(\omega,k^2) &=& E(k^2) - \frac{e g_{\pi N} \omega^2_c}{ 32 
\pi^2 m F_\pi^2} \biggl( 1 - \frac{5\omega_c}{ 2m}\biggr) \, \sqrt{1-y} 
\, \, ,  \\
L_{0+}(\omega,k^2) &=& L(k^2) - \frac{e g_{\pi N} \omega^2_c}{32 \pi^2
  m F_\pi^2 }\, \frac{1}{ 2 +\rho}  
\biggl( 1 - \frac{\omega_c}{ 2m} \frac{10+6\rho+\rho^2}{2+\rho}
\biggr) \, \sqrt{1-y} \, \, \, , \nonumber
\label{s0+app}
\end{eqnarray}
and the chiral expansion, Eq.(\ref{s0+gen}), is applied to ${\cal
  S}(k^2)$. This means that we can calculate ${\cal S} (k^2)$ at 
threshold in the isospin limit. The Born terms are readily evaluated,
\begin{eqnarray}
E(k^2)^{\rm Born} &=& \frac{e g_{\pi N}}{ 8 \pi m^2 } \biggl\{ -M_\pi
+\frac{1}{ 2m} \bigl[ (3+\kappa_p)M_\pi^2-(1+\kappa_p)k^2\bigr]
\nonumber \\ 
& & \qquad \quad + \frac{M_\pi}{ 8m^2} \bigl[ (5+4\kappa_p)k^2 
- 3(5+2\kappa_p)M_\pi^2\bigr] \biggr\}  \\
L(k^2)^{\rm Born} &=& \frac{e g_{\pi N} }{ 8 \pi m^2 } \biggl\{ -M_\pi
+\frac{3M_\pi^2-k^2}{ 2m} +\frac{M_\pi }{ 8 m^2} \bigl[ (5-2\kappa_p)k^2 -
15M_\pi^2\bigr] \biggr\} \nonumber \,  \, . 
\label{sborn}
\end{eqnarray}
The order $q^3$ one--loop contributions are given by 
\begin{eqnarray}
E(k^2)^{\rm q^3-loop} &=& \frac{e g_{\pi N} \omega_c^2}{128 \pi^2 m F_\pi^2}
\biggl[ \frac{\rho}{1+\rho} +\frac{(2+\rho)^2}{2 (1+\rho)^{3/2}}
\arccos \frac{-\rho}{2+\rho} \biggr] \, \, , \nonumber \\
L(k^2)^{\rm q^3-loop} &=& \frac{e g_{\pi N} \omega_c^2}{ 128 \pi^2 m F_\pi^2}
\biggl[ \frac{2}{1+\rho} +\frac{\rho}{(1+\rho)^{3/2}} 
\arccos \frac{-\rho}{ 2+\rho} \biggr] \, \, .
\label{s3loop}
\end{eqnarray}
These are finite and can be checked against the result of the
relativistic calculation presented in \cite{bklm}. To get an idea
about the uncertainties induced by the approximation on the energy
dependence discussed above, we have also evaluated to this order the
full ${\cal S} (\omega,k^2)$. The difference of the full result and the
approximation takes the form
\begin{eqnarray}
\Delta E_{0+}^{\rm q^3-loop} &=& 
\frac{e g_{\pi N} \omega_c^2}{128 \pi^2 m  F_\pi^3}
\biggl\{ \frac{\sqrt{y}}{y+\rho} \,
\biggl[ \rho - (2y +3\rho) \sqrt{1-y}+\frac{4y+4\rho+\rho^2}{2\sqrt{y+\rho}}
 \nonumber \\ & \times & \, H(y, \rho) \biggr] 
- \frac{\rho}{1+\rho} - \frac{(2+\rho)^2}{2(1+\rho)^{3/2}} \,
\arccos \frac{-\rho}{2+\rho} - 4 \sqrt{1-y} \biggr\} \, \, \, ,
\nonumber \\
\Delta L_{0+}^{\rm q^3-loop} &=& 
\frac{e g_{\pi N} \omega_c^2}{128 \pi^2 m  F_\pi^3}
\biggl\{ \frac{y^{3/2}}{y+\rho} \,
\biggl[ 2 - 2 \sqrt{1-y}+\frac{\rho}{\sqrt{y+\rho}}
\, H(y, \rho) \biggr]  \nonumber \\
&-& \frac{2}{1+\rho} - \frac{\rho}{(1+\rho)^{3/2}} \,
\arccos \frac{-\rho}{2+\rho} + \frac{4}{2+\rho} \sqrt{1-y} \biggr\} \, \, \, .
\label{s3fumapp}
\end{eqnarray}
We will come back to this later on. The order $q^4$ contributions to
${\cal S}(k^2)$ can be grouped into the graphs which are proportional
to $g_A$ and the ones proportional to $g_A^3$. These are no longer
finite and depend on the scale of dimensional regularization, $\lambda$. 
The counter terms to be discussed below have an infinite piece to
cancel these divergences and their scale--dependence is such that it
cancels the one from the $q^4$ loops. The explicit form of these terms
is (we first give the result of the $g_A^3$ graphs)
\begin{eqnarray}
E(k^2)^{\rm ga^3} & = &
\frac{e g_A^3 \omega_c^3}{ 128 \pi^3 m F_\pi^3}
\biggl\{\bigl(\frac{8}{3} + \frac{7\rho}{6} \bigr) \frac\ln{\omega_c}{\lambda}
-\frac{26}{9 } -\frac{59\rho}{36 } +\pi \bigl(\frac{5}{3} +\rho\bigr) 
\nonumber \\ & + &
\bigl(\frac{8}{3} + \frac{7\rho}{6} \bigr)\sqrt{1+\frac{4}{\rho}}\,
\ln \frac{\sqrt  {4+\rho}+\sqrt{\rho}}{ 2}  \\ 
&- &2(2+\rho) \int_0^1dx\sqrt{1-x^2+\rho
x(1-x)} \arccos \frac{x}{ \sqrt{1+\rho x(1-x)}} \biggr\} 
\nonumber \, \, \, ,
\label{e0+ga3}
\end{eqnarray}

\vspace{-1cm}

\begin{eqnarray}
L(k^2)^{\rm ga^3} & =& \frac{e g_A^3 \omega_c^3}{ 128 \pi^3 m F_\pi^3}
\biggl\{\bigl(\frac{2}{3} -\frac{5\rho}{ 6}\bigr) \ln \frac{\omega_c}{\lambda}
-\frac{8}{9} +\frac{13\rho}{36} +\pi \bigl(\frac{1}{6}
-\frac{\rho}{2}\bigr)
\nonumber \\ &  + &
\bigl(\frac{5}{3} + \frac{\rho}{6}\bigr)\sqrt{1+\frac{4}{\rho}} \,
\ln \frac{\sqrt {4+\rho}+\sqrt{\rho}}{2} \nonumber \\
& +&\int_0^1 dx
\frac{4(1+\rho)x^2+(\rho^2-2\rho-1)x-\rho(1+\rho)/2  -2}{\sqrt{1-x^2+\rho 
x(1-x)}}  \nonumber \\ & \times &
\arccos \frac{x}{\sqrt{1+\rho x(1-x)}} \biggr\} \, \, , 
\label{l0+ga3}
\end{eqnarray}
Similarly, one finds for the diagrams proportional to $g_A$ (these
form an independent subset of gauge invariant Feynman diagrams)
\begin{eqnarray}
E(k^2)^{\rm ga} & =& \frac{e g_A \omega_c^3}{128 \pi^3 m F_\pi^3}
\biggl\{\bigl(10+\frac{11\rho}{6}\bigr) \ln\frac{\omega_c}{\lambda}
-\frac{7}{3}  -\frac{31\rho}{36 } + \bigl(\frac{4}{3} + \frac{11\rho}{6} 
\bigr)\sqrt{1+\frac{4}{\rho}} \nonumber \\ 
& \times & \ln  \frac{\sqrt{4+\rho}+\sqrt{\rho}}{2}  -  
\frac{\pi\rho}{ 1+\rho} -\frac{\pi(2+\rho)^2}{2 (1+\rho)^{3/2}} 
\arccos \frac{-\rho}{2+\rho} \\ 
&+& \int_0^1dx \frac{ 4(1+\rho)x^2-(4+6\rho)x-4 }{ \sqrt{1-x^2
+\rho  x(1-x)}} \arcsin\frac{x}{\sqrt{1+\rho x(1-x)}} \biggr\} 
\, \, , \nonumber
\label{e0+ga}
\end{eqnarray}

\vspace{-1cm}

\begin{eqnarray}
L(k^2)^{\rm ga} &-& E(k^2)^{\rm ga}  = \frac{e g_A \omega_c^3(1+\rho)}{128\pi^3
m F_\pi^3} \biggl\{-2 \ln\frac{\omega_c}{\lambda}-3 \nonumber \\
&+&\int_0^1 dx \frac{x(2x-1)}{1-x^2 + \rho x(1-x)}
\biggl(-2-\rho + \frac{\pi \bigl(5 + 9\rho x / 2
-4(1+\rho)x^2\bigr)}{\sqrt{1-x^2+\rho x(1-x)}} \nonumber \\
& +& \frac{2+(\rho-2)x-2(1+ \rho)x^2}{ \sqrt{1-x^2+\rho x(1-x)}}
\arcsin \frac{x}{ \sqrt{1+\rho x(1-x)}}\biggr) \biggr\} \, \, \, .
\label{l0+ga}
\end{eqnarray}
Notice that the last integrand contains an integrable singularity  of the type
$1/\sqrt{1-x}$  for all $\rho>0$. As a check, the last integral in 
Eq.(\ref{l0+ga})  at $\rho=0$ gives $\pi^2 -3\pi+9$. The one--loop
corrections to order $q^4$ within the approximation on the
energy--dependence are thus determined. Finally, we turn to the
counter terms at this order. First, there are the two terms from
${\cal L}_{\pi N}^{(4)}$ which appeared already in the photoproduction
case. We have refitted the sum $a_1+a_2$ since in
Refs.\cite{bkmz,bkmprl} the full $\omega$--dependence of $E_{0+}$ was   
considered. Within the same approximation used here, the value of
$a_1+a_2$ is $7.85 \, \, {\rm GeV}^{-4}$ as compared to 
$a_1+a_2 = 6.60 \, \, {\rm GeV}^{-4}$ from the full
$\omega$--dependence. We will also use this latter value as a measure
for the theoretical uncertainty of our calculations. For the S--waves,
we have a priori two new counter terms at order $q^4$,
\begin{eqnarray}
E(k^2)^{\rm ct} &=& eM_\pi\bigl\{ (a_1+a_2)M_\pi^2 -a_3 k^2\bigr\} 
\, \, \, , \nonumber \\
L(k^2)^{\rm ct} & =& e M_\pi\bigl\{( a_1+a_2)M_\pi^2 -a_3 k^2 +
a_4(M_\pi^2-k^2)\bigr\} \, \, \, , 
\label{elct4}
\end{eqnarray}
so that $L(k^2) - E(k^2) \sim (1+\rho)$. However, as proven in the
appendix, in the soft pion limit one can show that
\begin{equation}
a_3 + a_4 = 0  \quad , \label{a34con}
\end{equation}
and thus there is only one LEC and furthermore a strong correlation 
between $E(k^2)^{\rm ct}$ and $L(k^2)^{\rm ct}$ at order $q^4$.
The low--energy constant $a_3$ will be treated as a free parameter and
pinned down by a fit to the available differential cross section data
at $k^2 = -0.1$~GeV$^2$. It turns out, however, that with a
$k^2$--independent $L^{\rm ct}(k^2)$, i.e. with the $k^2$--dependence
of $L(k^2)$ coming solely from the Born and loop graphs, one is not
able to fit the existing data. We therefore have to include the first
corrections to the soft--pion constraint Eq.(\ref{a34con}) away from 
the chiral limit. This induces terms of the type
\begin{equation}
E_{0+}^{\rm ct} , L_{0+}^{\rm ct} \sim a_5 \, M_\pi^2 \, k^2 \quad ,
\label{a5}
\end{equation}
which are arising from terms in the Lagrangian ${\cal L}_{\pi
  N}^{(5)}$ and are thus of higher order. These are the minimal terms
one has to take to be able to describe the data at $k^2 =
-0.1$~GeV$^2$. Of course, there are other counter terms at this order.
These, however, merely amount to quark mass renormalizations of the 
already considered $k^2$--independent
counter terms and will therefore be set to zero here.
In the next section, we show how to estimate the pertinent low--energy
constants from resonance exchange. A fully consistent ${\cal O}(q^5)$
calculation would also include two--loop graphs and one loop graphs
with insertions from ${\cal L}_{\pi N}^{(3)}$ (besides others). We
take here the pragmatic approach and subsume all these effects in the
effective couplings entering the higher order low--energy constants.
In addition, the form factor correction due to the finite proton Dirac
radius, cf. Eq.(\ref{F1v}), contributes as follows,
\begin{equation}
E(k^2)^{\rm rad} = L(k^2)^{\rm rad} = 
-\frac{e g_{\pi N} M_\pi k^2 }{48 \pi m^2 }\, \delta r_{1p}(\lambda) \, \, \, ,
\label{a3rad} 
\end{equation}
and is completely fixed from the knowledge of $<r^2>_1^V$ \cite{mmd},
$\delta r_{1p}(\lambda = m) = 6.60 \pm 0.29 \, \, {\rm GeV}^{-2}$
using $g_A$ as determined from the Goldberger--Treiman relation, $g_A
= g_{\pi N} F_\pi / m = 1.328$.\footnote{This value is not very
different from the one in the relativistic calculation, 
$\delta r_{1p}^{\rm rel}(\lambda = 1\, {\rm GeV}) = 7.35 
\, \, {\rm GeV}^{-2}$ \cite{bklm}.} This contribution  is
always treated separately.
We now turn to the estimate of the size of the resonance contributions
to the $a_i$ ($i=1,2,3,4$) and the P--wave low--energy constant $b_P$.


\section{Resonance saturation of the low--energy constants}

In this section, we consider the resonance saturation hypothesis to
estimate the numerical values of the various low--energy
constants. This principle works very well in the meson sector, see
e.g. refs.\cite{reso,reso1,reso2} and is also a good tool to estimate
the LECs in the presence of baryons, see e.g. \cite{bkmr} and \cite{ulfmit}. 
In the baryon sector, resonance saturation proceeds in two
steps. First, the effective field theory contains mesonic ($M$) and
baryonic $(N^*$) excitations chirally coupled to the Goldstone bosons
and the nucleons. Considering these excitations as very heavy, but
keeping the ratios of coupling constants to masses fixed, one produces
a string of higher dimensional operators involving only pions and
nucleons with coefficients given in terms of the resonance
parameters. Second, one then performs the heavy mass expansion for the
nucleons, i.e.
 \begin{equation}
\tilde{{\cal L}}_{\rm eff} [U, M , N, N^*] \to 
\bar{{\cal L}}_{\rm eff} [U, N] \to {\cal L}_{\rm eff} [U, N_v] \, \, \, ,
\label{chain}
\end{equation}
where $N_v$ denotes the velocity--dependent heavy nucleon field (in
general, we suppress the subscript '$v$').

First, consider t--channel vector meson exchange, in this case the
coupling of the $\rho^0$ and the $\omega$ to the nucleon and the
subsequent vector meson decay into the $\pi^0 \gamma^\star$ (a general
discussion concerning the coupling of spin--1 fields to the
pion--nucleon system has recently been given \cite{bm}). This leads to 
\begin{eqnarray}
a_1^V&+&a_2^V=a_3^V=-a^4_V = 
\frac{g_{\rho N} \, (1 +\kappa_\rho ) \,
  G_{\pi \rho \gamma}}{4 \, \pi \, m \, M^2_\rho } +
\frac{g_{\omega N} \, (1 +\kappa_\omega ) \,
  G_{\pi \omega \gamma}}{4 \, \pi \, m \, M^2_\omega } \, \, \, ,
\nonumber \\ b_P^V &=& 
\frac{g_{\rho N} \, G_{\pi \rho \gamma}}{2 \, \pi \,  M^2_\rho} +
\frac{g_{\omega N} \, G_{\pi \omega \gamma}}{2 \, \pi \,  M^2_\omega}
\, \, \, , \label{rhoomf}
\end{eqnarray}
where the $V \pi^0 \gamma$ couplings can be determined from the
radiative widths $\Gamma (V \to \pi^0 \gamma)$, $V = \rho^0, \omega$.
The tensor to vector coupling ratio $\kappa_V$ is known to be large
for the $\rho$, $\kappa_\rho = 6 \ldots 6.6$, small for the $\omega$,
$\kappa_\omega = -0.16 \pm 0.01$. The $\rho N$ coupling constant is
fairly well known whereas there is some sizeable uncertainty
concerning $g_{\omega N}$. To avoid these uncertainties, we use a
simpler form based in part on the gauged Wess--Zumino action, using
the KSFR relation and setting $\kappa_\rho = 6$, $\kappa_\omega = 0$
(for details, see \cite{bkmz}),
\begin{equation}
a_1^V+a_2^V =a_3^V=-a_4^V= \frac{1}{16 \pi^3 m F_\pi^3 } \, ,
\quad b_P^V = \frac{5}{(4\pi F_\pi)^3} \, \, \, , \label{aiV}
\end{equation}
leading to $a_1^V+a_2^V = 2.67 \, {\rm GeV}^{-4}$ and 
$b_P^V = 3.13 \, {\rm GeV}^{-3}$ (see also \cite{bkmz}).

From the s--channel (baryonic) excitations, the dominant one is
$\Delta (1232)$ exchange. We use the following $\Delta N \pi$ and $\Delta
N \gamma$ Lagrangians
\begin{eqnarray}
{\cal L}_{\Delta N\pi} &=& \frac{3 \sqrt{2}g_{\pi N}}{ 4m} \bar
\Delta^\mu \Theta_{\mu\nu}(Z) \partial^\nu \vec \pi \cdot \vec T^\dagger N +
{\rm h.c.} \nonumber \\
{\cal L}_{\Delta N\gamma} & =& \frac{ig_1}{2m} \bar \Delta^\mu 
\Theta_{\mu\lambda}(Y) \gamma_\nu\gamma_5 F^{\nu\lambda} T^{3\dagger}
N  \nonumber \\ & -&
\frac{g_2}{4m^2} \bar \Delta^\mu \Theta_{\mu\nu}(X) \gamma_5 F^{\nu\lambda}
T^{3\dagger}\partial_\lambda N \nonumber \\ 
& -& \frac{g_3 }{ 4m^2} \bar \Delta^\mu 
\Theta_{\mu\nu}(X') \gamma_5 (\partial_\lambda F^{\nu\lambda}) T^{3\dagger} N +
{\rm h.c.} 
\, \, \, , \label{Ldelta}
\end{eqnarray}
with
\begin{equation}
\Theta_{\mu\nu}(Z) =g_{\mu\nu}-\bigl(Z+{1\over2}\bigr)
\gamma_\mu\gamma_\nu \quad . \label{thetamunu}
\end{equation}
$N$ and $\Delta_\mu$ denote the relativistic spin--1/2 (Dirac) and
spin--3/2 (Rarita--Schwinger) field, respectively.
The third term in ${\cal L}_{\Delta N\gamma}$ vanishes for real
photons \cite{schol}. The off--shell parameters $X$, $Y$ and $Z$ are
severely constrained from $\pi N$ scattering, the nucleon
polarizabilities and the fit to the new TAPS data for neutral pion 
photoproduction, for details see Ref.\cite{bkmprl}. Using
Eqs.(\ref{Ldelta}), one finds for the $\Delta$ contribution to the $a_i$
(to order $q^4$ in the Lagrangian)
\begin{eqnarray}
a_1^\Delta &+& a_2^\Delta = \frac{g_{\pi N}\sqrt{2}}{ 24 \pi m^3
m^2_\Delta} \biggl\{ g_1\biggl( \frac{m_\Delta^2 - m \, m_\Delta /2 -m^2}{
m_\Delta - m}  + m Y(3-2Z)  \nonumber \\
& &  \qquad \qquad \qquad \qquad \qquad 
+  m_\Delta(Y+Z+4YZ)\biggr) \nonumber \\ 
& & \quad + g_2\biggl(\frac{3m}{8}
(2X+1)(1-2Z)+ \frac{m_\Delta}{4}(1+X+Z+4XZ)\biggr)\biggr\} \nonumber \\
a_3^\Delta &=& -a_4^\Delta = a_1^\Delta +a_2^\Delta  + \frac{g_{\pi N}
\sqrt{2}}{ 96 \pi m^3 m^2_\Delta} 
\biggl\{ m(2Z-1)\bigl[8 g_1 Y +g_2(2X+1)\bigr] \nonumber \\ 
& & \quad + g_3
\bigl[ m(2X'+1)(2Z-1) - 2m_\Delta(1+X'+Z+4X'Z)\bigr]\biggr\} 
\, \, \, . \label{aidelta}
\end{eqnarray}
As discussed in the previous section, we have to account for the first
correction to the soft--pion theorem $a_3+a_4=0$. This is entirely
given by $\Delta$--exchange and contributes to $L_{0+}$ as 
follows,\footnote{We remark that this contribution to $L_{0+}$ is the
one which really gives the first correction to the soft-pion theorem
whereas the equivalent $E_{0+}$ term is a quark--mass renormalization
of $a_3$.}
\begin{equation}
L_{0+}^{{\rm ct}, q^5} = - e \,  M_\pi \, k^2 \,  (a_3 +a_4)^\Delta
\quad , 
\end{equation}
with
\begin{eqnarray}
(a_3 +a_4)^\Delta &=&  \frac{g_{\pi N} \sqrt{2} M_\pi}{ 48 \pi m^3 m^2_\Delta} 
\biggl\{ g_1 \biggl[2 Y (4Z-1) - \frac{m_\Delta}{m_\Delta-m} \biggr] \nonumber
\\& +& g_2\biggl[ \frac{X}{2}(4Z-1) + \frac{3m_\Delta^2-4m_\Delta m
  +3m^2}{4 m (m_\Delta - m)} \,\biggr] \nonumber \\ 
& & + g_3 \biggl[Z-X' - \frac{m_\Delta}{m}(1+X'+Z+4X'Z)\biggr]\biggr\} 
\, \, \, , \label{aq5l}
\end{eqnarray}
and similarly for $E_{0+}$
\begin{eqnarray}
a_3^\Delta = \frac{g_{\pi N} \sqrt{2} M_\pi}{48  \pi m^3 m_\Delta^2}
& \times & \nonumber \\
 \biggl\{  g_1 \biggl[ -m^2  (4YZ&+&4Y+Z+\frac{3}{2}) - mm_\Delta
(8YZ-4Y+2Z+\frac{1}{2}) \nonumber \\ 
 + m^2_\Delta(28YZ&-&4Y-7Z-7) -
\frac{4m^3_\Delta}{m}(4YZ+Y+Z+1) \biggr]\frac{1}{(m-m_\Delta)^2}
\nonumber \\
+ g_2 \biggl[ m (18YZ&-&7Y+9Z-\frac{7}{2}) - m_\Delta
(30XZ-4X+12Z+\frac{1}{2}) \nonumber \\ & & \quad \qquad\quad \qquad + 
\frac{m^2_\Delta}{m}(12XZ+3X+3Z
+\frac{1}{2} ) \biggr]\frac{4}{m-m_\Delta} \nonumber \\
+ g_3 \biggl[ -8X'Z&-&4Z+4X'+2 + \frac{4m_\Delta}{m} (1+ 4X' Z + X'
+Z) \biggr] \biggr\}
\label{aq5e}
\end{eqnarray}
A good check on the lengthy expressions Eqs.(\ref{aq5l},\ref{aq5e})
(together with the analogous terms of order $M_\pi^4$ not shown here)
is that they fulfill the constraint $L(k^2) - E(k^2) \sim (1+\rho)$.

Furthermore, $b_P^\Delta$ is the same as in the photoproduction case
\cite{bkmz},
\begin{eqnarray}
b_P^\Delta = \frac{g_{\pi N}g_1\sqrt{2}}{12 \pi m^2
  m^2_\Delta} & \biggl\{ &
\frac{2 m_\Delta^2 +m_\Delta m-m^2 }{ 2(m_\Delta-m)} \nonumber \\
& + & m( Y+Z+2YZ) +m_\Delta (Y+Z+4YZ)\biggr\}
\, \, \, . \label{bpdelta}
\end{eqnarray}
The couplings $g_1$, $g_2$ and the off--shell parameter $X,Y,Z$ have
been previously determined from neutral pion photoproduction, $\pi N$
scattering and the nucleons' electromagnetic polarizabilities. Nothing
is known about the signs and magnitudes of $g_3$ and $X'$. These only
show up in reactions with virtual photons. In what follows, we will
use these two parameters to obtain a best fit to the MAMI and NIKHEF
data at $k^2 = -0.1$~GeV$^2$. Since we do not constrain the magnitudes
of $g_3$ and $X'$, this effectively amounts to a free fit with $a_3
\neq a_4$. We finally remark that $g_3$ and $X'$ also enter virtual
Compton scattering and can eventually be pinned down within some 
broad ranges in the future.


\section{Fit to the existing data}

In this section, we show the combined fit to the NIKHEF 
($\epsilon = 0.67$) \cite{benno} and the
MAMI ($\epsilon = 0.582$ and $0.885$) \cite{dist} data at $k^2 = -0.1$ 
GeV$^2$ and the resulting multipoles. Note that the MAMI data are 
$\phi$--integrated differential cross sections involving only $R_T$
and $R_L$. In Figs.2a,b, the dashed lines give the best fit at order $q^4$,
i.e. with the soft--pion 
constraint $a_3 + a_4 = 0$. Clearly, this does not describe the data.
If one adds, however, the order $q^5$ counter terms for the S--wave
multipoles as discussed in the previous sections, one finds an
acceptable fit (solid lines) with a $\chi^2$/dof of 2.296. In the
resonance exchange picture, this amounts to 
\begin{equation}
g_3 = -125.7 \pm 6.2 \,\,\, , \qquad X' =-0.22 \pm 0.09 \, \, \, \, . 
\label{fitval}
\end{equation} 
Clearly, such a large value of the unknown third $N \Delta
\gamma$ coupling constant indicates that one subsumes in its value
other effects like e.g. from two--loop graphs. Alternatively, 
one can only take the $q^4$ counter terms and relax the
constraint $a_3 = -a_4$ in the fit. This  leads to 
$a_3 = -1.37$~GeV$^{-4}$ and $a_4 = -0.22$~GeV$^{-4}$, which are
numbers of natural size. We remark that in a combined fit to the
NIKHEF and the MAMI data, it is not possible to correctly get the
normalization of the two MAMI data sets for the two different values
of the photon polarization $\epsilon$, compare Fig.2b.

\hskip .6in
\epsfysize=5.5in
\epsffile{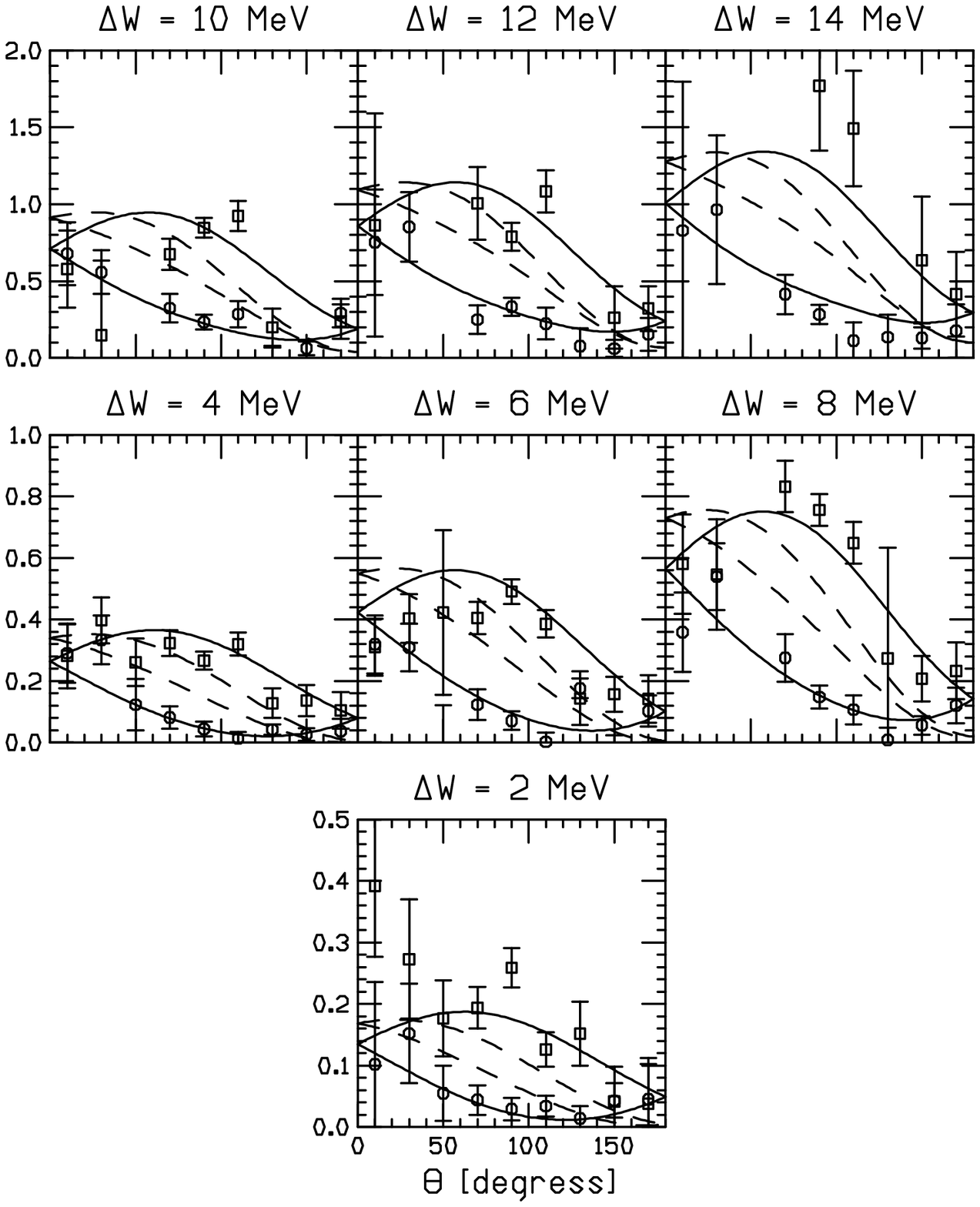}

\vspace{0.3cm}


{\centerline{Fig. 2a: NIKHEF data.
Open squares (circles): $\phi = 180^\circ \, (0^\circ)$. }}


\hskip 3.2in
\epsfysize=3.2in
\epsffile{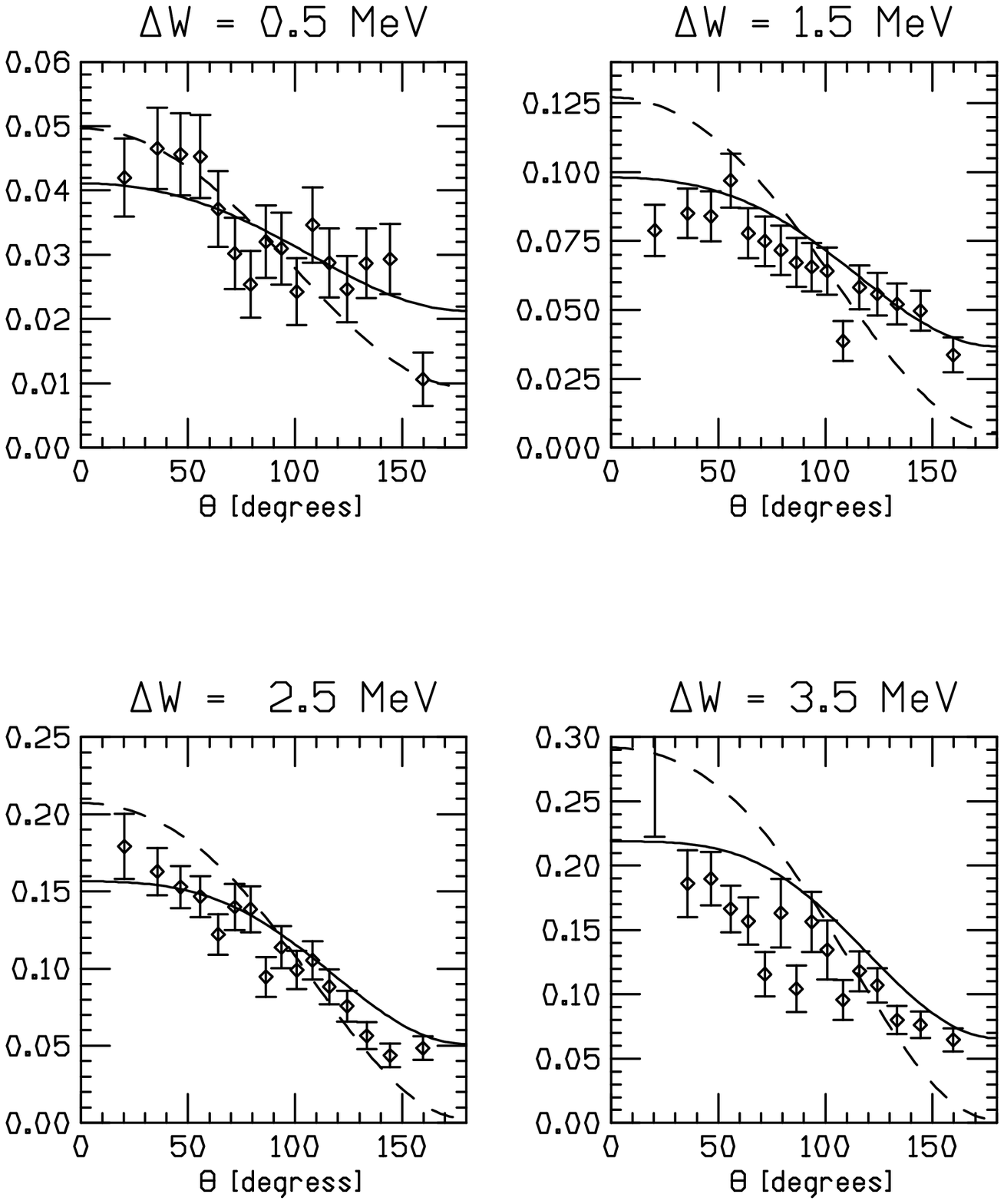}

\vspace{-3.2in}

\epsfysize=3.2in
\epsffile{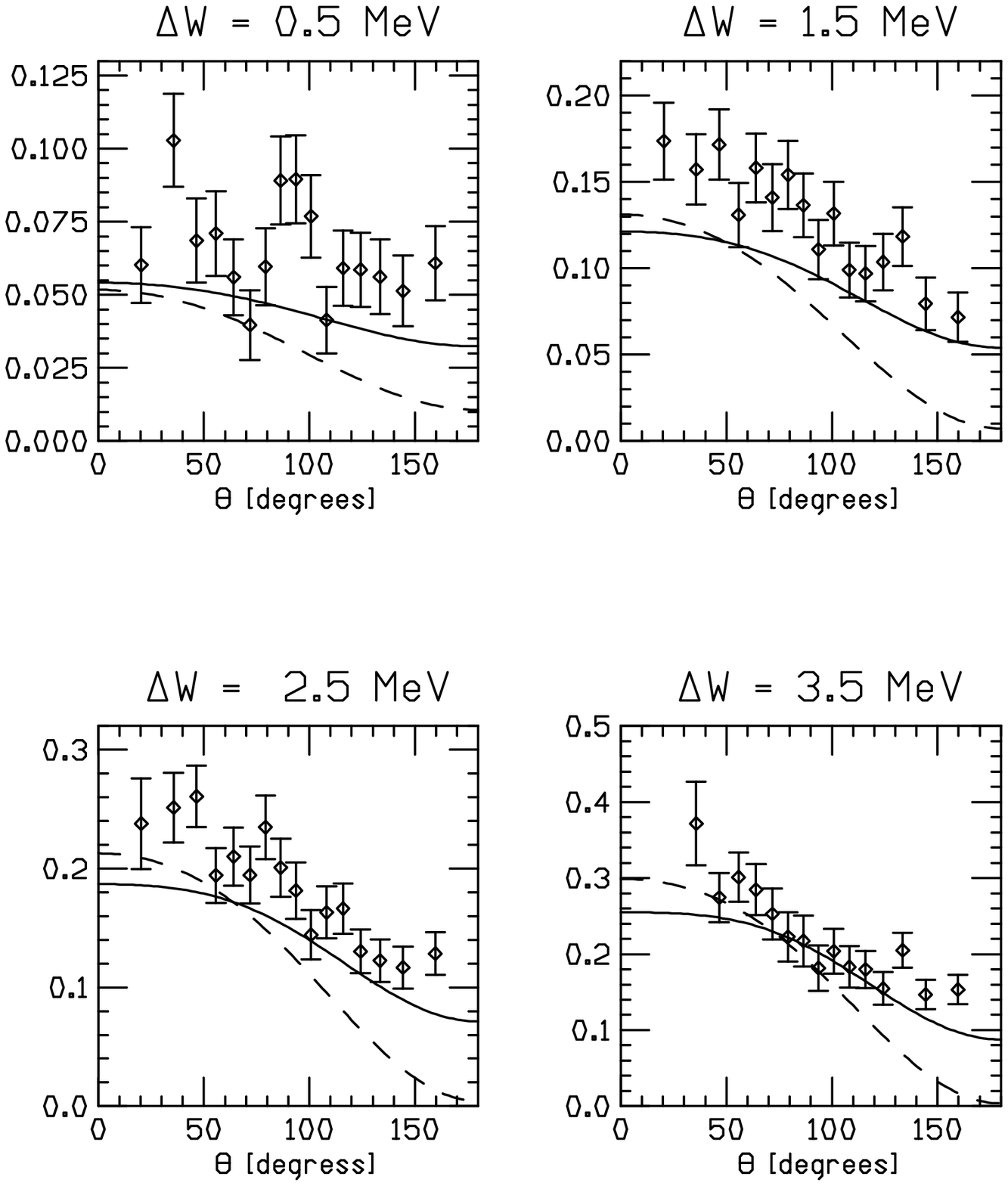}

\vspace{0.3cm}

{\centerline{Fig. 2b: MAMI data. Left (right) panel: $\epsilon = 0.885
    \, (0.582)$. }}

The corresponding multipoles are shown in Fig.3 for $\Delta W = 0
\ldots 15$~MeV, all in units of $10^{-3}/M_{\pi^+}$. For the S--waves, we
give the real and the imaginary parts, indicated by the solid and
dashed lines, respectively. We remark that Re~$E_{0+}$ has changed
sign as compared to the photoproduction case, it shows the typical
cusp effect at the opening of the $\pi^+n$ threshold. In contrast,
Re~$L_{0+}$

\hskip 1.in
\epsfysize=3.8in
\epsffile{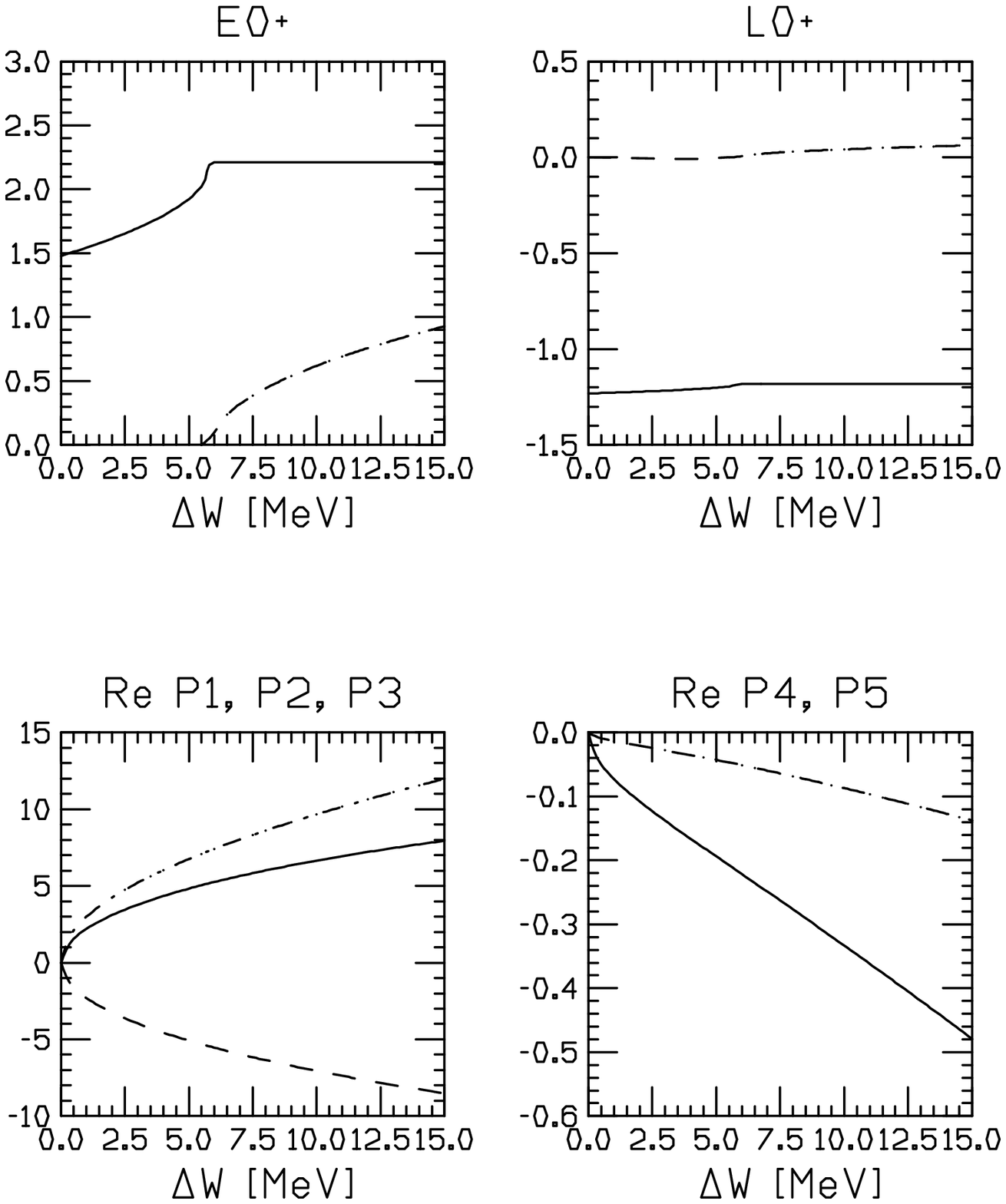}

\vspace{0.1cm}

\centerline{Fig. 3: Multipoles at $k^2 = -0.1$~GeV$^2$. 
                    See text for notations.}

is essentially energy--independent with a very small cusp. 
For the P--waves, we only show the real
parts since the imaginary parts are very small. The large P--waves are
all of the same size as shown in the lower left panel ($P_1,P_2,P_3$:
solid, dashed, dash--dotted line, in order) whereas $P_4$ and $P_5$ are more
than one magnitude smaller, see the lower right panel in Fig.3
($P_4,P_5$: solid and dashed line, respectively).

In Fig.4, we show the effect of the approximation on the
$\omega$--dependence. Without refitting any parameters, we have added
the correction Eq.(\ref{s3loop}) to the S--wave multipoles. 
The solid curves in Fig.4 refer to the approximation and the
dash--dotted ones to the complete $\omega$--dependence in the loops to
order $q^3$. We see
that the effect on $E_{0+}$ is small and somewhat more pronounced in
$L_{0+}$. We remark that these curves should only be considered
indicative since there is also a difference in the $q^4$--loops, which
we did not evaluate in detail. Furthermore, part of this effect would
be absorbed in the values of the LECs which we used in the fit.

\hskip 1.in
\epsfysize=2in
\epsffile{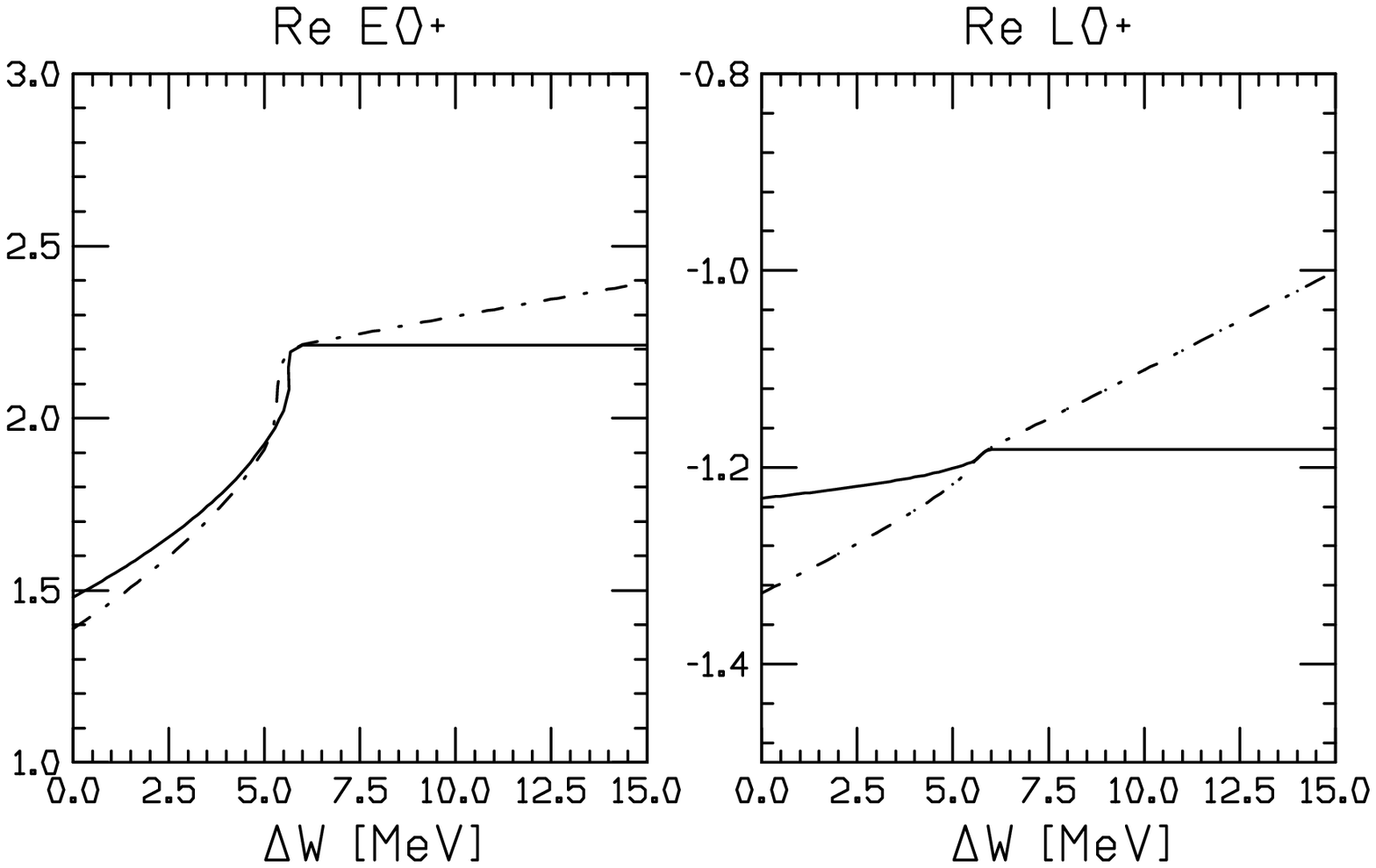}

\vspace{0.1cm}

\centerline{Fig. 4: Approximation on the $\omega$--dependence at order
  $q^3$.}

Furthermore, we have also performed some fits with the larger value of
$b_p = 15.8$ GeV$^{-3}$. It leads to a somewhat larger (in magnitude)
value of $g_3 = -145.8 \pm 6.2$ and $X' = -0.26 \pm 0.06$, consistent
with the value given in Eq.(\ref{fitval}). The $\chi^2$/dof is 2.945,
i.e. worse than for the smaller value of $b_p = 13.0$~GeV$^{-3}$. To
improve the fits with the larger $b_P$, one thus would have to
readjust the parameters in the photoproduction case. Clearly, this
discrepancy deserves further experimental clarification. In what
follows, we will always use the smaller value of $b_p$. We would like
to stress again, see also \cite{bklm}, that for a good test of chiral 
dynamics, one needs data at lower photon virtualities as witnessed by
the large value we find for the coupling constant $g_3$ and the
necessity of using a particular dimension five operator. However, as $|k^2|$
decreases, the latter becomes less and less important. Finally, we
note that the convergence in the S--wave loops and counter terms is
slow, similar to the photoproduction case \cite{bkmz}, whereas the
Born terms at order $q^5$ can safely be neglected.
 
\newpage

\section{Predictions at lower photon virtualities}

Having fixed all parameters at $k^2 =-0.1$ GeV$^2$, we can now make
predictions for lower photon virtualities. We will concentrate here
on the range of $|k^2|$ between 0.04 and 0.06 GeV$^2$ since at the
upper end some data have been taken at MAMI and it is conceivable that
in the near future one will not be able to go below 0.04 GeV$^2$ since
the pion has to leave the target.\footnote{Predictions at lower photon
virtualities can be supplied upon request.}

In Figs.~5 and 6, we show the real parts of the S--wave and the
large P--wave multipoles as a function of $\Delta W$ for $k^2 = -0.06,
-0.05, -0.04$ GeV$^2$ (solid, dashed--dotted and dashed lines, in
order). At threshold, Re~$E_{0+}$ passes zero for $k^2 \simeq -0.04$ GeV$^2$.

\hskip 0.8in
\epsfysize=2.2in
\epsffile{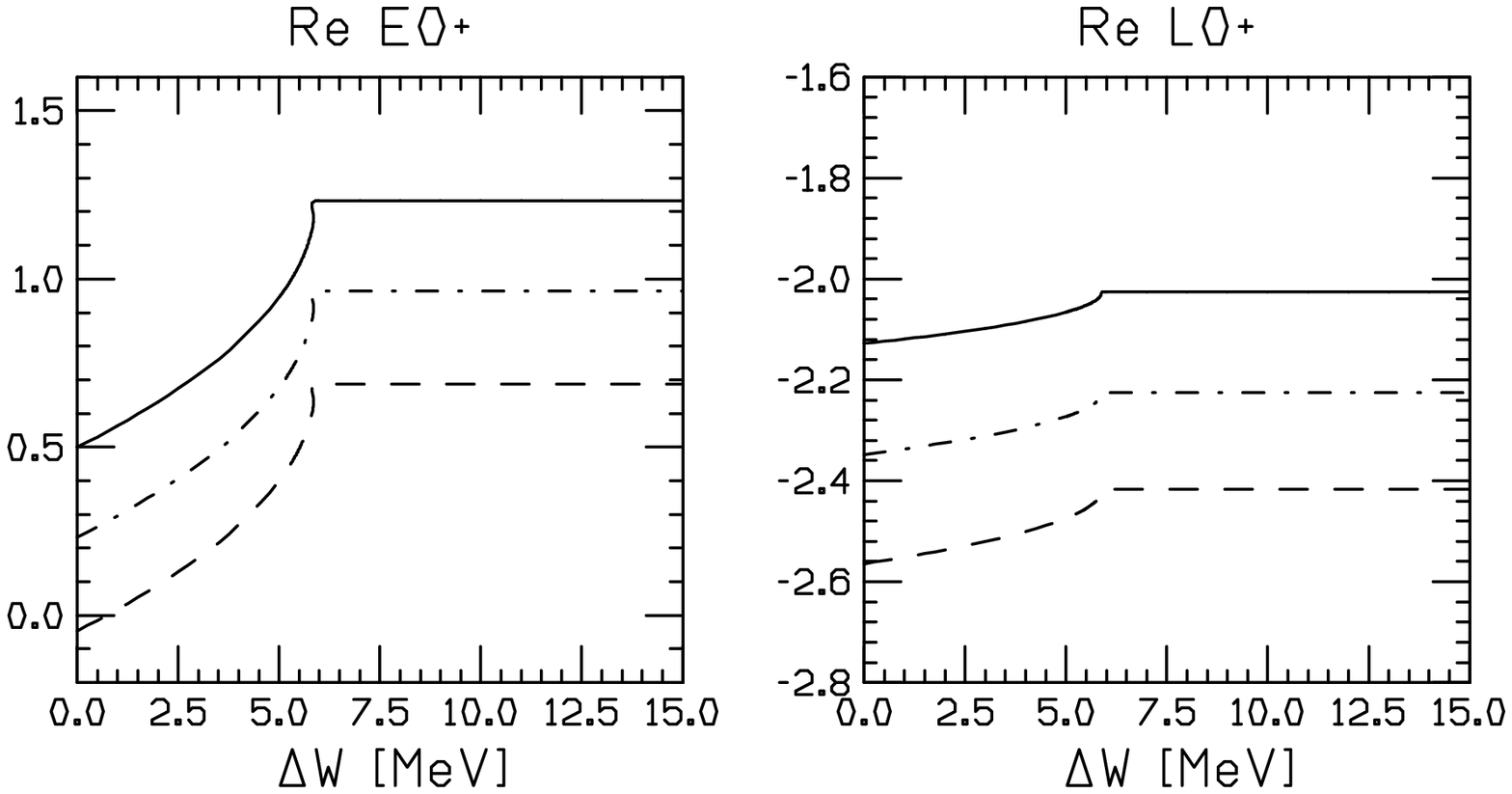}

\vspace{0.1cm}

\centerline{Fig. 5: Predictions for the S--wave mulitpoles. For
  notations, see text.}

\medskip \bigskip

\epsfysize=2.3in
\epsffile{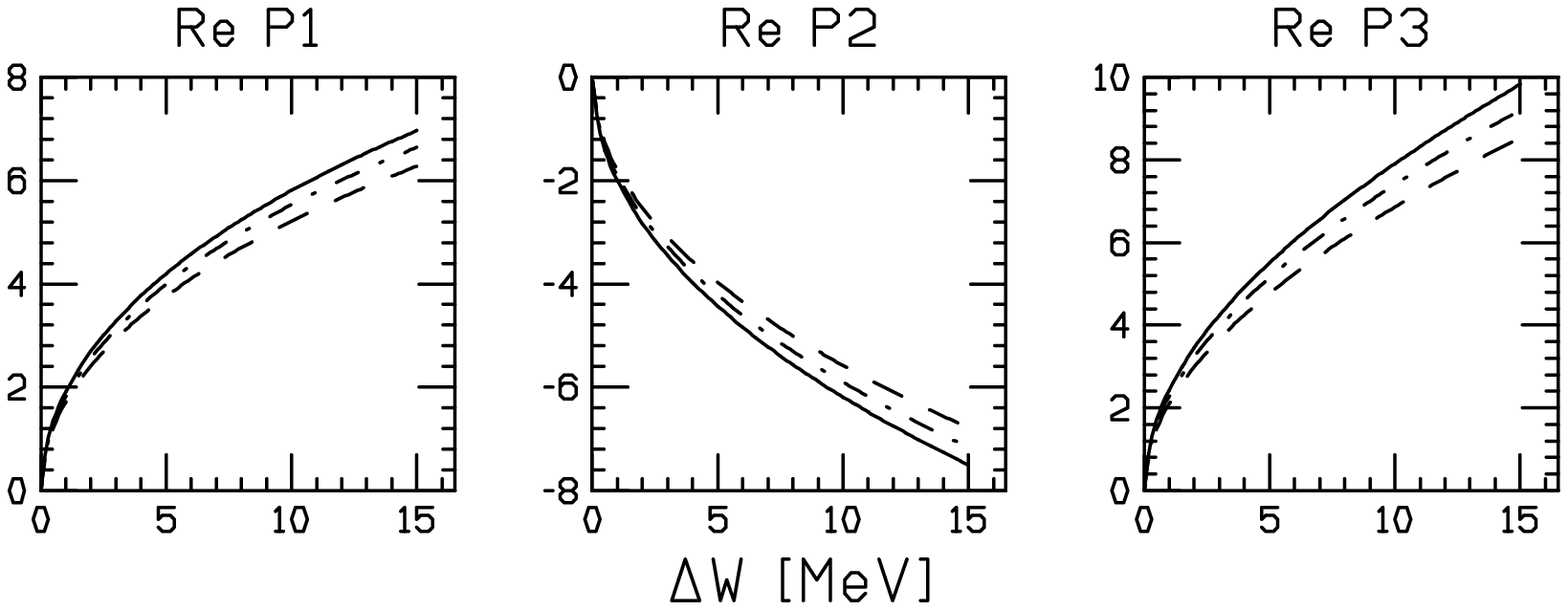}

\vspace{0.1cm}

\centerline{Fig. 6: Predictions for the P--wave mulitpoles. For
  notations, see text.}
  
For the new MAMI data, we show the various differential cross sections
at $k^2 = -0.06$ GeV$^2$ with $\epsilon = 0.582$ and $\Delta W = 2$ and $8$
MeV as indicated by the solid and dashed--dotted lines, respectively.

\bigskip

\hskip 1.in
\epsfysize=3.5in
\epsffile{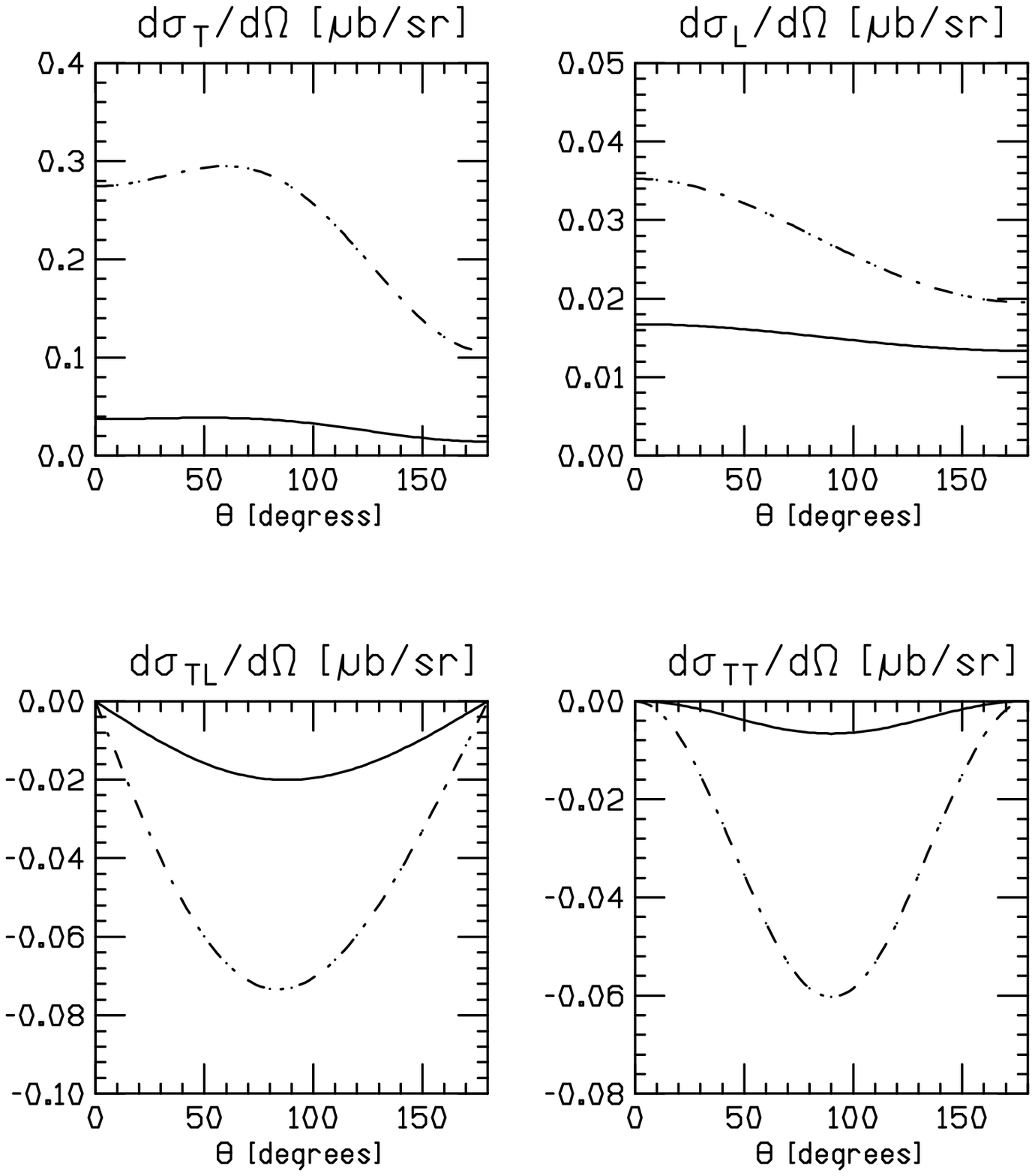}

\vspace{0.1cm}

\centerline{Fig. 7: Predictions for the differential cross sections. For
  notations, see text.}

The S--wave cross section $a_0$, defined as
\begin{equation}
a_0 = |E_{0+}|^2 + \epsilon_L \, |L_{0+}|^2 \, \, \, \, ,
\label{a0}
\end{equation}
is shown in Fig.8.  The data of Welch et al. \cite{pat} (open squares)  are
taken at different values of $\epsilon$. This range is indicated by
the dotted lines in Fig.8. The solid line refers to $\epsilon = 0.67$,

\hskip .8in
\epsfysize=2.5in
\epsffile{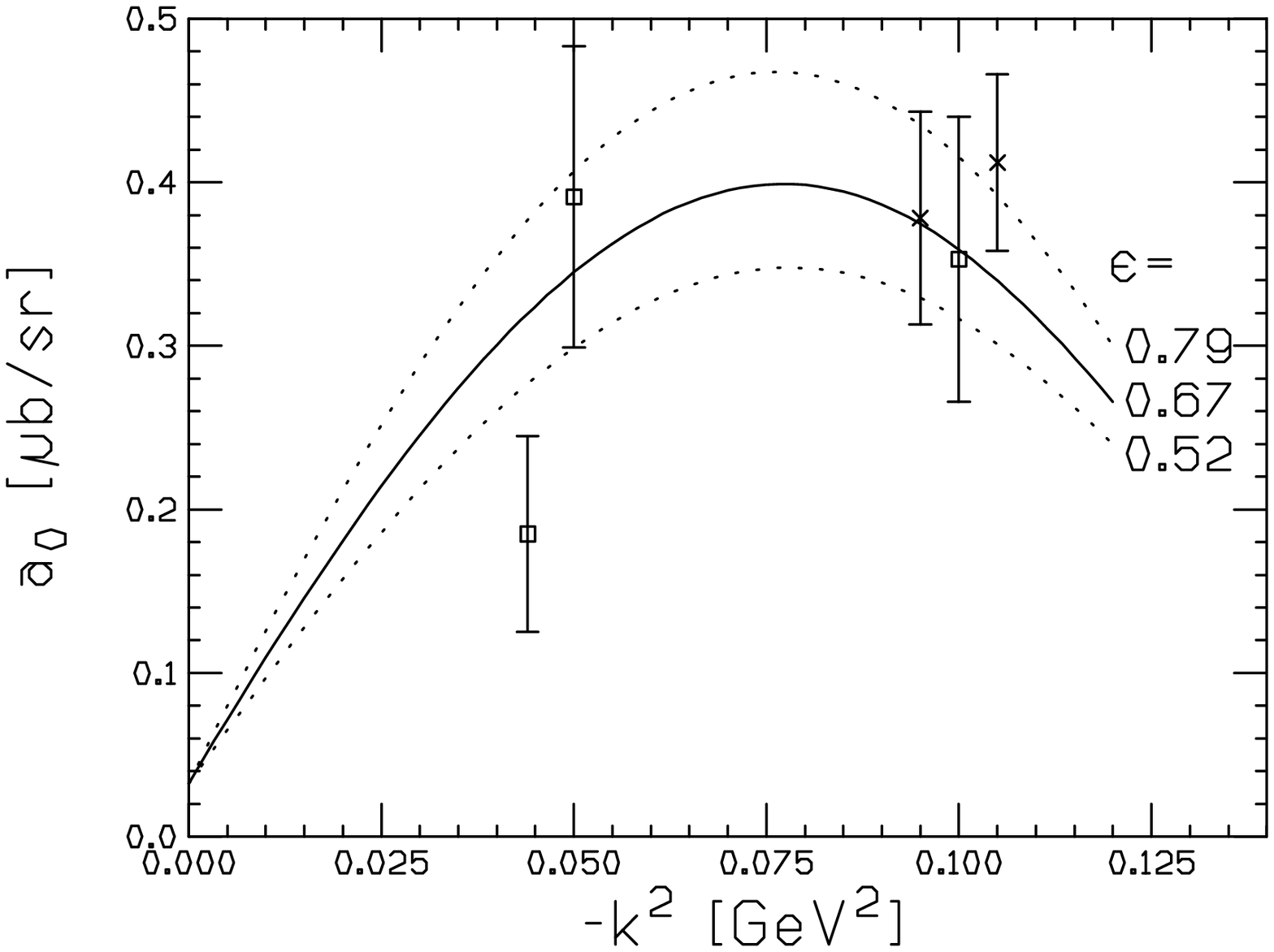}

\vspace{0.1cm}

\centerline{Fig. 8: S--wave cross section. For notations, see text.}
i.e. the kinematics of van den Brink et al. \cite{benno} (crosses). 
Within the sizeable uncertainty, the S--wave cross section $a_0$ 
as predicted by the calculation is
consistent with the data and shows again the flattening with
increasing $|k^2|$ as already found in the relativistic 
calculation \cite{bklmprl}.


\section{Summary and outlook}

In this paper, we have considered neutral pion electroproduction
off protons in the framework of heavy baryon chiral perturbation
theory. The main results can be summarized as follows:
\begin{enumerate} 
\item[$\bullet$]Extending the successful photoproduction calculation
\cite{bkmz} \cite{bkmprl} to account for all $k^2$--dependent counter
terms at order $q^4$ does not give a decent fit to the existing 
electroproduction data at the rather large $k^2 = -0.1$~GeV$^2 \simeq
-5 M_\pi^2$.  This can be traced
back to a particular soft--pion theorem which severely constrains the
possible local dimension four contact terms. 
We are thus forced to take into account a
particular dimension five operator which  gives the first correction
to the soft--pion theorem. With that term, the data can be fitted.
\item[$\bullet$]We have given predictions for the $|k^2|$ range  between
0.04 and 0.06 GeV$^2$, where new data have been and are going to be
taken. At  threshold, the real part of $E_{0+}$ changes sign at $k^2
\simeq -0.04$ GeV$^2$. The calculated S--wave cross section is in
satisfactory agreement with existing determinations. More detailed
predictions pertaining to the relevant kinematics of future
experiments are available from the authors.
\end{enumerate}

Further improvements of the calculations presented here are:
\begin{enumerate} 
\item[$\bullet$]The P--wave multipoles should be calculated to one
order higher. This would give a more accurate description of the small
P--waves and can further be used to tighten the predictions of
the P--wave LETs \cite{bkmlet}.
\item[$\bullet$]The full--energy dependence of the loops should be
taken into account when more accurate data will become available. 
\item[$\bullet$]A more systematic study of higher order effects than
done here should be done to eventually pin down the $N\Delta \gamma$ 
parameters $g_3$ and $X'$.
\end{enumerate}

Clearly, when more and more accurate data at lower photon virtualities
will be available, neutral pion electroprodcution offers a variety of
tests of the chiral QCD dynamics.


\section{Acknowledgements}

We are grateful to J.C. Bergstrom, A.M. Bernstein, H. Blok,
M. Distler, D. Drechsel, O. Hanstein, L. Tiator,
H.B. van den Brink and Th. Walcher for useful
comments and communications.


\appendix
\section{Counterterms at order $q^4$}
\def\theequation{\Alph{section}.\arabic{equation}}
\setcounter{equation}{0}
\label{appa}

Here, we wish to proof Eq.(\ref{a34con}) which states that there are
no polynomial S--wave counter terms with $a_3 + a_4 \ne 0$
up-to-and-including order $q^4$. For that, we follow appendix E of 
Ref.\cite{bklm} and consider the soft--pion limit for the process
$\gamma^\star (k) + p (p_1) \to \pi^0 (q) +p (p_2)$, i.e. 
for $q_\mu \to 0$ (this
includes the chiral limit $M_\pi \to 0$). In this limit, only the
nucleon pole graphs (including the electromagnetic form factors and,
in particular, the radius counter term) 
remain, for all values of the photon virtuality $k^2$. The
corresponding non--pole amplitude (denoted by an 'overbar') must
thus vanish,
\begin{eqnarray}
\overline{J}_\mu = i \, \bar{u}_2 \, \gamma_5 \, \biggl[ \,
\gamma_\mu \, \bar{B}_5 &+&  2P_\mu \, (\bar{B}_1 + \bar{B}_2
-m\bar{B}_6 ) \nonumber \\
&+&  k_\mu \, ( \bar{B}_1 + 2\bar{B}_4 -
2 m \bar{B}_7) \biggr] \, u_1 = 0
\label{Tbar}
\end{eqnarray}
where we use the notation of Ref.\cite{bklm}, 
i.e. $P = (p_1+p_2)/2$ (see section 2 of that review).
In the soft pion limit, the following conditions arise
\begin{equation}
\bar{B}_5 =0 \, \, \quad \bar{A}_1+\bar{B}_2 = 0\, \, , \quad
\bar{B}_1+2 \bar{B}_4 -2 m \bar{A}_6 = 0 \, \, \, .
\label{ABcon}
\end{equation}
In fact, the $\bar{B}_5$, $\bar{B}_1+2 \bar{B}_4$ and $\bar{A}_6$ are
individually zero since they are odd under crossing, i.e. proportional
to $(s-u) \sim q \cdot (p_1+p_2)$.

Modulo irrelevant prefactors and setting $m =1$, the S--wave
multipoles are related to the invariant amplitudes $A_i$ and $B_i$ as
follows (we drop the subscript '$0+$' and refer to \cite{bklm} for
definitions)
\begin{eqnarray}
E & =& \mu \bar{A}_1 + \mu^2 \bar{A}_3 + \frac{\mu}{2} (\mu^2 - \nu)\bar{A}_4
- \nu \bar{A}_6  \nonumber \\
L - E &=& \frac{1}{2} (\mu^2 - \nu ) [ -\bar{A}_1 -\bar{B}_2 + \bar{B}_1 
+ 2\bar{B}_4 - \mu \bar{A}_4 - 2 \bar{A}_6 \, ] \, \, ,
\label{ELAB}
\end{eqnarray}
with $\nu \sim k^2 \sim \rho$ and $\mu \sim M_\pi$.
To order $q^4$, the polynomial terms (in $k^2$) contributing to $L$,
taking into account the crossing properties of the various amplitudes, 
are solely given by  $L = \mu \, \bar{A}_1 - \nu \, 
(\bar{B}_1 + 2\bar{B}_4)/2$,  with 
\begin{equation}
1) \, \, \, \bar{A}_1 = \alpha \, k^2 \, \, \, \, {\rm and} \, \, \, \,
2) \, \, \, \bar{B}_1+2\bar{B}_4 = \beta \, (s-u)/2 = 2 \, \beta \, \mu \,
\, \, .
\label{a34pol}
\end{equation}
Two cases are possible that could lead to $a_3+ a_4 \ne  0$, namely 
\begin{equation}
i) \, \, \, \alpha \, \, {\rm or} \, \, \beta \ne 0 \, \, , \quad
ii) \, \, \,  \alpha \, \, {\rm and} \, \, \beta \ne 0 \, \,  
{\rm with} \, \, \alpha \ne \beta \, \, .
\label{cases}
\end{equation}
However, gauge invariance ($\overline{J} \cdot k = 0)$ (see Eq.(2.1)
of Ref.\cite{bklm}) implies that
$\bar{B}_2 = -\beta \, k^2$. Furthermore,  the second condition
from Eqs.(\ref{ABcon}) leads to $\bar{B}_2 = -\alpha \, k^2$ (because of
1) in Eq.(\ref{a34pol})). Clearly, $\alpha$ and $\beta$ must be equal
and both possibilities in Eq.(\ref{cases}) are ruled out. This gives
the desired result, $a_3 + a_4 = 0$. 

To summarize the argument, the soft pion theorem
excludes a polynomial of the type $\bar{A}_1 = {\rm const} \, k^2$
which would give rise to $a_3 + a_4 \ne 0$ for the non--pole
contributions. Only the case 
$\bar{A}_1 = - \bar{B}_2 = {\rm const} \, k^2 \, , \, \, \bar{B}_{14} 
= {\rm const} \, (s-u)/2$ is allowed leading immediately to the
constraint.  Evidently, this constraint does not apply to the nucleon
pole graphs, compare Eq.(\ref{a3rad}).

\vspace{1cm}

\end{document}